\documentstyle[12pt]{article}

\begin{document}

\newcommand{\inner}[2]{\langle #1\mid #2\rangle}
\newcommand{\bra}[1]{\langle #1\mid}
\newcommand{\ket}[1]{\mid #1\rangle}
\newcommand{\Ft}{{\cal F}}
\newcommand{\Ct}{{\cal C}}
\newcommand{\St}{{\cal S}}
\newcommand{\Tt}{{\cal T}}
\newcommand{\Kt}{{\cal K}}
\newcommand{\Qt}{{\cal Q}}
\newcommand{\DD}{{\cal D}}
\newcommand{\HH}{{\cal H}}

\newcommand{\JJ}{{\cal J}}

\newcommand{\LL}{{\cal L}}

\newcommand{\square}{\vrule height 1.5ex width 1.2ex depth -.1ex }

\newcommand{\II}{{\rm I}}
\newcommand\CC{\mkern1mu\raise2.2pt\hbox{$\scriptscriptstyle|$}
                {\mkern-7mu\rm C}}
\newcommand{\RR}{{\rm I\! R}}
\newcommand{\SS}{{\rm S}}
\newcommand{\Coinfd}{C_0^\infty(\RR^d\backslash\{ 0\})}
\newcommand{\Coinf}[1]{C_0^\infty(\RR^{#1}\backslash\{ 0\})}
\newcommand{\domin}{\dot{\ominus}}

\newtheorem{Thm}{Theorem}[section]
\newtheorem{Def}[Thm]{Definition}
\newtheorem{Lem}[Thm]{Lemma}
\newtheorem{Prop}[Thm]{Proposition}
\newtheorem{Cor}[Thm]{Corollary}
\renewcommand{\theequation}{\thesection.\arabic{equation}}
\newcommand{\sect}[1]{\section{#1}\setcounter{equation}{0}}

\newcommand{\Hr}{H_{\rm res}}
\newcommand{\hr}{h_{\rm res}}
\newcommand{\hrp}{h_{\rm res,+}}
\begin{titlepage}
\renewcommand{\thefootnote}{\fnsymbol{footnote}}

\rightline{DAMTP-R93/21}
\vspace{0.5in}
\LARGE
\center{Resonance Point Interactions}
\Large
\vspace{0.5in}
\center{C.J. Fewster\footnote{E-mail address:
C.J.Fewster@amtp.cam.ac.uk}} \vspace{0.2in}
\large
\center{\em Department of Applied Mathematics and Theoretical Physics,
\\ University of Cambridge,
\\  Silver Street, Cambridge CB3 9EW, U.K.}
\vspace{0.2in}
\center{\today}

\vspace{0.5in}

\begin{abstract}
A new construction is presented for point interactions (PI) and
generalised point interactions (GPI). The construction is an inverse
scattering procedure, using integral transforms suggested by the
required scattering theory. The usual class of PI in 3 dimensions
(i.e. the self adjoint extensions of the Laplacian on the domain of
smooth functions compactly supported away from the origin) is
reconstructed. In addition a 1-parameter family of GPI models termed
resonance point interactions (RPI) is constructed, labelled by $M$.
The case $M<0$ coincides with a special case of a known GPI model;
the case $M>0$ appears to be new. In both cases, the Hilbert space of
states must be extended, for $M<0$, a larger Hilbert space is
required, whilst for $M>0$, the Hilbert space is extended to a
Pontryagin space. In the latter case, the space of physical states is
identified as a positive definite invariant subspace. Complete
M{\o}ller wave operators are constructed for the models considered,
using a two space formalism where necessary, which confirm that the
PI and RPI models exhibit the required scattering theory. The
physical interpretation of RPI as models for quantum mechanical
systems exhibiting zero energy resonances is described.
\end{abstract}

\setcounter{footnote}{0}
\renewcommand{\thefootnote}{\arabic{footnote}}
\end{titlepage}

\sect{Introduction and Main Ideas}

Point interactions (PI) have long been of interest as solvable models
in quantum mechanics (see \cite{Alb} for an extensive bibliography).
Heuristically, they represent Hamiltonians with $\delta$-function
potentials ``$H=-\triangle + \lambda \delta(x)$'', although it is
well known that such Hamiltonians fail to make rigorous sense in
dimensions $d\ge 2$. Instead, a point interaction situated at the
origin is rigorously defined as one of the self-adjoint extensions of
the Laplacian on $\Coinfd$, i.e. smooth functions compactly supported
away from the origin. One of the most useful features of PI is that
they represent a leading order approximation to the scattering theory
of non-point interactions. For example, in 3 dimensions, the low
energy expansion of the $S$-wave partial wave shift $\delta_0(k)$
\cite{Newt} is
\begin{equation} \cot\delta_0(k) = -\frac{1}{kL} +
r_0k + O(k^3), \label{eq:low}
\end{equation}
where $L$ and $r_0$ are the $S$-wave scattering length and effective
range respectively, and the expansion is valid for spherically
symmetric potentials  decaying at least exponentially at infinity,
and for $L\not=0,\infty$. In $d=3$,  there is a 1-parameter family of
PI, which may be labelled $\{H^L\mid L\in\RR\cup\{\infty\}\}$, whose
scattering theory is non-trivial only in the $S$-wave, where it is
given by
\begin{equation}
\cot\delta_0(k) = -\frac{1}{kL}, \label{eq:PIlow}
\end{equation}
thus approximating~(\ref{eq:low}) at leading order.

The PI described so far suffer from various limitations. Firstly, as
noted for example by Grossmann and Wu \cite{GWu}, they are restricted
only to dimensions $d\le 3$, because $-\triangle$ is essentially
self-adjoint on $\Coinfd$ in dimensions 4 and higher. Secondly, for
$d=2,3$, they yield non-trivial scattering only in the sector of zero
angular momentum. Thirdly, they give at best only the leading order
approximation to the scattering theory, and in the special cases
$L=0,\infty$ fail to do even that, for the phase shift for a
potential with scattering length $L=0,\infty$ does not obey
$\cot\delta_0(k)=O(k^{-1})$ at leading order, but rather
\begin{equation}
\cot\delta_0(k)=pk^{-3} +O(k^{-1})
\end{equation}
for $L=0$, and
\begin{equation}
\cot\delta_0(k) = qk +O(k^3) \label{eq:linf}
\end{equation}
for $L=\infty$.
To remove these limitations on PI, various generalised point
interactions (GPI) have been proposed, which correspond heuristically
to Hamiltonians with $\delta$-derivative and more general
distributional potentials. Models of this type were first discussed
by Shirokov \cite{Shiro} and were given a mathematical foundation by
Pavlov \cite{Pav1,Pav2,Pav3} and Shondin \cite{Shond1,Shond2} (see
also \cite{Diejen}). GPI are not defined on the usual Hilbert space
$L^2(\RR^3)$, but on a larger space: either an extended Hilbert space
$L^2(\RR^3)\oplus \CC^n$ \cite{Shiro,Shond1,Pav1,Pav2,Pav3}, or a
Pontryagin space\footnote{A Pontryagin space $\Pi$ is an inner
product space which admits a direct orthogonal decomposition
$\Pi=\HH_+\ominus\HH_-$ into a Hilbert space $\HH_+$ with positive
definite inner product, and a finite dimensional Hilbert space
$\HH_-$ with negative definite inner product. The dimension of
$\HH_-$ is called the rank of indefiniteness. See \cite{Bognar}.}
$\Pi = L^2(\RR^3)\oplus \CC^m\ominus \CC^m$
\cite{Shiro,Shond2,Diejen}. Due to the presence of negative and zero
normed states, Pontryagin spaces do not immediately admit the usual
probability interpretation of quantum mechanics. However, for GPI
models it is often the case that there is a positive definite
subspace $\HH_+$ of $\Pi$ which is invariant under the unitary
evolution generated by the GPI Hamiltonian and therefore has a
natural interpretation as the space of physical states.

In conventional treatments of GPI, one decides at the outset how the
Hilbert space is to be extended, and then constructs the class of GPI
which `live' on this space, either by a generalisation of the von
Neumann theory of deficiency indices \cite{Pav3}, or by constructing
separable perturbations of the free Laplacian in accordance with a
generalised version of Krein's formula \cite{Shond2,Diejen,Pav2}. In
this paper, we present a new construction of point and generalised
point interactions, which removes the necessity to determine the
underlying inner product structure in advance. Rather, the
appropriate extension to the initial space $L^2(\RR^3)$ emerges
naturally in the course of the construction.

Our construction depends on the observation that for a point-like
`potential', the scattering phase shifts completely determine the
generalised eigenfunctions of the continuum spectrum. To see this,
note that for a  potential compactly supported within radius $a$ of
the origin, the phase shifts determine the eigenfunctions (up to
normalisation) for $r>a$, and therefore if $a\rightarrow 0$, the
phase shift in a given angular momentum sector determines the
continuum eigenfunctions in that sector for all values of the radial
coordinate.\footnote{In contrast to the usual treatments of GPI's on
Pontryagin spaces, we neglect the possibility of distributional
contributions at the origin. See the discussion in Section 7.} Thus,
for example, in the $S$-wave, we have the radial eigenfunctions
$u_k(r)$ at wavenumber $k$
\begin{equation}
u_k(r) = \left(\frac{2}{\pi}\right)^{1/2} \sin (kr+\delta_0(k)).
\label{eq:gefn}
\end{equation}

We introduce the notation $\HH_r=L^2((0,\infty),dr)$ and
$\HH_k=L^2((0,\infty),dk)$ for Hilbert spaces of square integrable
functions on position and momentum spaces respectively. Armed with
the generalised eigenfunctions $u_k$, one can define an integral
transform $U:\HH_r\rightarrow\HH_k$ so that
\begin{equation}
f(r) = (U^{-1}\tilde{f})(r) = \left(\frac{2}{\pi}\right)^{1/2}
\int_0^\infty \sin (kr+\delta_0(k))\tilde{f}(k) dk .\label{eq:sph}
\end{equation}
If $U$ were unitary, then one could immediately define a self-adjoint
operator $H=U^*k^2 U$ for which the $u_k(r)$ would constitute a
complete set of generalised eigenfunctions: one would thereby have
constructed the spectral representation of $H$. In general, the
mapping $U$ is not unitary and so one cannot proceed in this way. For
relatively simple functions $\delta_0(k)$, it is, however, possible
to determine exactly how $U$ fails to be unitary. One can then extend
either $\HH_r$ or $\HH_k$ (or both) in such a way that $U$ may be
extended to a unitary mapping $\hat{U}$. Thus a self-adjoint operator
$H=\hat{U}^*k^2\hat{U}$ may be constructed, whose generalised
continuum eigenfunctions (projected onto the original position
Hilbert space $\HH_r$) coincide with the $u_k(r)$.

In particular, in the cases studied, it is easy to determine how the
Hilbert space should be extended. There is a certain amount of
ambiguity in our procedure (as in other constructions of GPI);
however, with the additional requirement of locality, it is possible
to make natural choices for the various free parameters occuring. We
also emphasise that our aim is not to construct the most general
possible class of GPI, but rather to construct at least one GPI with
the required scattering behaviour.

Although certain GPI models have previously been constructed by
inverse methods \cite{Shond1} the analysis usually proceeds
immediately from the $T$-matrix to the discussion of a candidate
resolvent $R(z)$ whose free parameters are then constrained by the
requirement that $R(z)$ actually be the resolvent of a self-adjoint
operator on the pre-determined inner product space. In contrast, the
present treatment focusses on integral transforms suggested by the
scattering data.

After some preliminaries in Section 2, we illustrate our construction
in two cases. Firstly, in Section 3, we construct the class of point
interactions $H^L$ with scattering theory given by~(\ref{eq:PIlow})
thereby reconstructing the self-adjoint extensions of $-\triangle$ on
$\Coinf{3}$, as expected. As is well known, for $L>0$, the
self-adjoint extension $H^L$ possesses a single bound state: in the
context of our construction, this is manifested in the need to extend
$\HH_k$. For $L<0$, there is no bound state, and neither $\HH_r$ nor
$\HH_k$ need be extended.

Secondly, in Section 4, we study the class of point interactions on
$\RR^3$ modelling potentials exhibiting zero energy resonances
(infinite scattering length). As noted above, the leading order
scattering behaviour of such systems~(\ref{eq:linf}) is not
well-approximated by any of the usual PI. Instead, we construct
generalised point interactions $\Hr^{M}$ ($M\in\RR\cup\{\infty\}$)
with scattering theory
\begin{equation}
\cot\delta_0(k) = kM   \label{eq:RPIlow}
\end{equation}
in the $S$-wave, and trivial scattering for higher angular momenta.
We refer to this class of GPI as {\em resonance point interactions}
(RPI). The special cases $M=0,\infty$ are identified with the PI
$H^\infty,H^0$ respectively. Apart from this, there are two cases:
$M>0$ and $M<0$. For $M<0$, we find that $\HH_r$ is extended to
$L^2((0,\infty),dr)\oplus \CC$, although the momentum Hilbert space
is unchanged. This case represents a subset of the models of type
$B_2$ discussed by Shondin \cite{Shond1}. For $M>0$, both $\HH_r$ and
$\HH_k$ are extended to Pontryagin spaces of form
$L^2(0,\infty)\ominus \CC$. This model appears to be new: moreover,
it is not presently clear how the usual GPI constructions could be
used to reproduce this model.

In Section 5, we verify that the PI and RPI models exhibit the
required scattering theory by explicitly contructing the M{\o}ller
wave operators (in a two space setting, where necessary). In Section
6, we consider the physical interpretation of  RPI models. To do
this, we employ a general methodology for discussing the `large scale
effects of small objects' developed by Kay and the author \cite{KF}.
In particular, we develop a {\em fitting formula} (analogous to those
given in \cite{KF}) for matching a given potential $V(r)$ with a zero
energy resonance to the `best fit' RPI. This leads to a natural
interpretation of the `extra dimension' in the extended Hilbert space
of the $M<0$ RPI models as representing a meta-stable state: part of
a wavepacket incident on such an RPI disappears from the Hilbert
space $L^2(\RR^3)$ and the `missing' probability is stored in the
extra dimension before being slowly released with a $O(t^{-1/2})$
time dependence.

It might be objected that neither case $L=0,\infty$ is generic, and
that, to all intents and purposes, the usual PI suffice to describe
the leading order behaviour. However, the motivation for the present
work arose in a consideration of the scattering of charged particles
off magnetic flux tubes of small radius \cite{FK}, in which it was
found that the scattering lengths for spin-$\frac{1}{2}$ particles
generically take the values $0$ or $\infty$ in certain angular
momentum sectors. In consequence, the conventional point interactions
(in this case the self-adjoint extensions of the operator describing
the dynamics in the background of an infinitesimally thin wire of
flux) fail to describe the leading order scattering theory in these
sectors. The special nature of this system can be attributed to the
fact that it is an example of supersymmetric quantum mechanics (see,
for example \cite{Thall}). Elsewhere \cite{F}, we will construct the
appropriate class of RPI for this system.

\sect{Preliminaries}

We develop various standard properties of the sine and cosine
transforms $\St$ and $\Ct$ and also demonstrate density of certain
subspaces which will be needed in the sequel. $\St$ and $\Ct$ are
defined by
\begin{eqnarray}
(\St f)(k) =\sqrt{\frac{2}{\pi}}\int_0^\infty dr \sin kr f(r) & &
(\Ct f)(k) =\sqrt{\frac{2}{\pi}}\int_0^\infty dr \cos kr f(r)
\end{eqnarray}
(the integrals are intended in the sense of `limit in the mean').
Both are unitary maps from $\HH_r$ to $\HH_k$, with inverses
\begin{eqnarray}
(\St^{-1} f)(r) =\sqrt{\frac{2}{\pi}}\int_0^\infty dk \sin kr f(k) & &
(\Ct^{-1} f)(r) =\sqrt{\frac{2}{\pi}}\int_0^\infty dk \cos kr f(k).
\end{eqnarray}
Define the operators $E_\pm:L^2(0,\infty)\rightarrow L^2(\RR)$ by
\begin{equation}
(E_\pm\varphi)(x) = \left\{\begin{array}{cl} \varphi(x) & x>0 \\
                                          \pm\varphi(x) & x<0.
\end{array}\right.
\end{equation}
$E_\pm\varphi$ are the even and odd continuations of $\varphi$
respectively. For clarity, we also define the restriction operator
$\JJ:L^2(\RR)\rightarrow L^2(0,\infty)$ by $(\JJ\varphi)(x) =
\varphi(x)$ for all $x\ge 0$. We also define Fourier transformation,
$\Ft$ by
\begin{equation}
(\Ft f)(k) = \int_{-\infty}^\infty \frac{dr}{\sqrt{2\pi}} e^{ikr}f(r).
\end{equation}
These operators allow us to express $\St$ and $\Ct$ in terms of
$\Ft$:
\begin{eqnarray}
(\Ct f)(k) = (\JJ\Ft E_+f)(k) & & (E_+\Ct f)(k)=(\Ft E_+f)(k)
\nonumber \\
(\St f)(k) = (\JJ\Ft E_-f)(k) & & (E_-\St f)(k)=(\Ft E_-f)(k).
\end{eqnarray}
These relationships entail that $\St$ and $\Ct$ possess many
properties inherited from Fourier transformation. The following is a
simple corollary of the Paley-Wiener theorem.
\begin{Prop} \label{Prop:PW}
Let $f\in C_0^\infty(0,\infty)\subset \HH_r$. Then $(\Ct f)(k)$
($(\St f)(k)$) is an even (odd), entire analytic function of $k$,
whose restriction to the real axis decreases faster than polynomially
at infinity. If $g\in C_0^\infty(0,\infty) \subset \HH_k$, then
$(\Ct^{-1} g)(r)$ ($(\St^{-1} g)(r)$) is an even (odd), entire
analytic function of $r$, whose restriction to the real axis
decreases faster than polynomially at infinity.
\end{Prop}

Next, we define the subspace $\DD^L$ of $L^2(\RR,dk)$ for
$L\in\RR\backslash\{0\}$ by
\begin{equation}
\DD^L = \{ (1+(kL)^2)^{1/2}\Ft f\mid f\in \Coinf{}\}
\end{equation}
and the normalised vector $\psi_L\in\HH$ by
\begin{equation}
\psi_L(k) = \left(\frac{2|L|}{\pi}\right)^{1/2} (1+(kL)^2)^{-1/2}.
\end{equation}
We then prove the following:
\begin{Lem} \label{Lem:dns}
$\DD^L\oplus\{\lambda E_+\psi_L\mid\lambda\in\CC\}$ is dense in
$L^2(\RR,dk)$ for any $L\in\RR\backslash\{0\}$.
\end{Lem}
{\em Proof:} Suppose $\varphi\in L^2(\RR)$ is orthogonal to $\DD$.
Then $(1+(kL)^2)^{1/2}\varphi$ is the Fourier transform of a
distribution supported at the origin. Hence
\begin{equation}
\varphi = \sum_{\alpha<n}\frac{a_\alpha k^\alpha}{(1+(kL)^2)^{1/2}}
\end{equation}
for some $a_\alpha, n$. Square integrability then forces $\varphi$ to
be a multiple of $E_+\psi_M$. $\square$

We now define the subspaces $\DD^L_\pm$ of $\HH_k$ by
\begin{eqnarray}
\DD^L_+ &=& \{ (1+(kL)^2)^{1/2}\Ct f\mid f\in C_0^\infty(0,\infty)\}
\nonumber \\
\DD^L_- &=& \{ (1+(kL)^2)^{1/2}\St f\mid f\in C_0^\infty(0,\infty)\}
\label{eq:Dpm}
\end{eqnarray}
\begin{Prop} \label{Prop:dense}
$\DD^L_+\oplus \{\lambda\psi_L\mid\lambda\in\CC\}$ and
$\DD^L_-$ are each dense in $\HH_k$.
\end{Prop}
{\em Proof:} Note that $\Coinf{}=E_+ C_0^\infty(0,\infty)\oplus E_-
C_0^\infty(0,\infty)$ and hence that $\DD^L=E_+\DD^L_+\oplus
E_-\DD^L_-$. By Lemma~\ref{Lem:dns}, $\DD^L\oplus\{\lambda
E_+\psi_M\mid\lambda\in\CC\}$ is a dense subspace of $L^2(\RR,dk)$
and admits an orthogonal decomposition into a subspace of even
functions and a subspace of odd functions. It is trivial to show that
the restrictions of these subspaces to $\RR^+$ must therefore be
individually dense in $\HH_k$, and the result follows. $\square$

Finally, the following well known identities are valid for all
$f(r)\in C_0^\infty(0,\infty)$
\begin{eqnarray}
k(\St f)(k) = \left(\Ct \frac{d}{dr}f\right)(k); & &
 k(\Ct f)(k) =- \left(\St \frac{d}{dr}f\right)(k) \label{eq:id1}
\end{eqnarray}
For $f(k)\in C_0^\infty(0,\infty)$, we have
\begin{eqnarray}
(\St^{-1} kf)(r) = -\frac{d}{dr} (\Ct^{-1}f)(r); & &
(\Ct^{-1} kf)(r) = \frac{d}{dr} (\St^{-1}f)(r) \label{eq:id2}
\end{eqnarray}

\sect{Point Interactions}

In this section, we put into practice the construction sketched in
Section 1 by reconstructing the familiar class of PI in 3 dimensions.
Our starting point is the scattering theory described
by~(\ref{eq:PIlow}) in the $S$-wave for some
$L\in\RR\cup\{\infty\}$, and $\delta_\ell(k)=0$ for all $k$ and all
$\ell\ge 1$. We therefore restrict attention to the $S$-wave and
attempt to make rigorous the heuristic spectral representation given
by~(\ref{eq:sph}), extending the various Hilbert spaces involved if
necessary.

In terms of $\St$ and $\Ct$, we define the bounded mapping
$\Ft_L:\HH_r\rightarrow\HH_k$ by
\begin{equation}
\Ft_L = (1+(kL)^2)^{-1/2}\St - kL(1+(kL)^2)^{-1/2}\Ct
\end{equation}
in terms of which equation~(\ref{eq:sph}) may be re-written as
\begin{equation}
f(r) = (\Ft_L^* \tilde{f})(r).
\end{equation}
In the special cases $L=0,\infty$, $\Ft_L$ reduces to $\St$ and $\Ct$
respectively. Restricting to $L\not=0,\infty$, we now compute
$\Ft_L^*\Ft_L$. We have
\begin{eqnarray}
\Ft_L^*\Ft_L &=& \II - \Ct^{-1}\frac{1}{1+(kL)^2}\Ct +
\St^{-1}\frac{1}{1+(kL)^2}\St \nonumber \\
 & & - \Ct^{-1}\frac{kL}{1+(kL)^2}\St -
\St^{-1}\frac{kL}{1+(kL)^2}\Ct .
\end{eqnarray}
Writing $\Ct^{-1}\frac{1}{1+(kL)^2}\Ct -
\St^{-1}\frac{1}{1+(kL)^2}\St$ as an integral kernel, we see that
\begin{eqnarray}
\frac{2}{\pi}\int_0^\infty dk
\frac{\cos kr\cos kr^\prime -\sin kr\sin kr^\prime}{1+(kL)^2} & = &
\frac{1}{2\pi}\int_{-\infty}^\infty
\frac{e^{ik(r+r^\prime)} +e^{-ik(r+r^\prime)}}{1+(kL)^2} \nonumber \\
&=& \frac{1}{|L|}e^{-(r+r^\prime)/|L|}
\end{eqnarray}
and so we find
\begin{equation}
\Ct^{-1}\frac{1}{1+(kL)^2}\Ct - \St^{-1}\frac{1}{1+(kL)^2}\St =
\frac{1}{2}\ket{\chi_L}\bra{\chi_L},
\end{equation}
where $\chi_L$ is given by
\begin{equation}
\chi_L(r)= \left(\frac{2}{|L|}\right)^{1/2}
\exp \left(-r/|L|\right). \label{eq:chi}
\end{equation}
Similarly,
\begin{equation}
\Ct^{-1}\frac{kL}{1+(kL)^2}\St + \St^{-1}\frac{kL}{1+(kL)^2}\Ct =
\frac{{\rm sgn} L}{2}\ket{\chi_L}\bra{\chi_L}.
\end{equation}

Hence, we deduce
\begin{equation}
\Ft_L^*\Ft_L =
\left\{\begin{array}{cl} \II - \ket{\chi_L}\bra{\chi_L} & L>0 \\
                                    \II & L <0. \end{array}\right.
\label{eq:FstarF}
\end{equation}
We now consider $\Ft_L\Ft_L^*$ and find
\begin{eqnarray}
\Ft_L\Ft_L^* & = &
\II -\frac{1}{(1+(kL)^2)^{1/2}}\St\Ct^{-1}
\frac{kL}{(1+(kL)^2)^{1/2}}\nonumber \\
&&- \frac{kL}{(1+(kL)^2)^{1/2}}\Ct\St^{-1}
\frac{1}{(1+(kL)^2)^{1/2}}.
\end{eqnarray}
On the domain $\DD^L_-$ defined in~(\ref{eq:Dpm}), the last term may
be re-written:
\begin{eqnarray}
\frac{kL}{(1+(kL)^2)^{1/2}}\Ct\St^{-1}\frac{1}{(1+(kL)^2)^{1/2}} & =&
\frac{L}{(1+(kL)^2)^{1/2}}
\St\frac{d}{dr}\St^{-1}\frac{1}{(1+(kL)^2)^{1/2}}\nonumber \\
 &= & \frac{1}{(1+(kL)^2)^{1/2}}
\St\Ct^{-1}\frac{kL}{(1+(kL)^2)^{1/2}}
\end{eqnarray}
using the identities~(\ref{eq:id1}) and~(\ref{eq:id2}). Thus we have
$\Ft_L\Ft_L^*=\II$ on $\DD^L_-$, which is dense by
Proposition~\ref{Prop:dense}. Hence
\begin{equation}
\Ft_L\Ft_L^* = \II. \label{eq:FFstar}
\end{equation}

Including the special cases $L=0,\infty$, we have thus proved
\begin{Prop}
For $L\le 0$ or $L=\infty$, $\Ft_L$ is a unitary mapping from
$\HH_r$ to $\HH_k$.
\end{Prop}

In the case $L>0$, $\Ft_L$ fails to be unitary, as it has non-trivial
kernel: $\Ft_L\ket{\chi_L}=0$. Note that this deficiency is
restricted to a 1-dimensional subspace. We can compensate for the
`missing probability' by extending $\HH_k$ to $\HH_k\oplus\CC$ (with
inner product
$\inner{f\oplus\alpha}{g\oplus\beta}=
\inner{f}{g}_{L^2}+\overline{\alpha}\beta$)
and then defining the mapping $\hat{\Ft_L}$ by
\begin{eqnarray}
\hat{\Ft_L} = \left[ \Ft_L, \bra{\chi_L}\right]:
\HH_r &\longrightarrow & \HH_k\oplus \CC \nonumber \\
f &\longrightarrow & \Ft_L f \oplus \inner{\chi_L}{f}
\end{eqnarray}
We then have
\begin{Prop}
For $L>0$, $\hat{\Ft_L}$ is a unitary mapping from
$\HH_r$ to $\HH_k\oplus \CC$, with inverse
\begin{equation}
\hat{\Ft_L}^{-1} (f\oplus\alpha) = \Ft_L^*f + \alpha\ket{\chi_L}.
\end{equation}
\end{Prop}
{\em Proof:} Note that $\Ft_L$ is a surjection onto $\HH_k$ as a
consequence of~(\ref{eq:FFstar}). The result then follows immediately
from the definition of $\hat{\Ft_L}$ and~(\ref{eq:FstarF}).
$\square$

Our extension to the momentum Hilbert space will, of course, carry
the interpretation of a bound state with normalised eigenfunction
$\chi_L(r)\in\HH_r$. Note that the mapping $\hat{\Ft}_L$ is just one
of a 1-parameter family of physically equivalent  possible mappings
$\hat{\Ft}_L^\delta = \left[ \Ft_L,e^{i\delta}\bra{\chi_L}\right]$
corresponding to re-phasing $\chi_L$.

We may now proceed to define the point interaction Hamiltonians. In
the case $L<0$, and also in the special cases $L=0,\infty$ we have
seen that the mapping $\Ft_L$ is unitary, and so we may immediately
define a self adjoint operator by
\begin{equation}
h^L = \Ft_L^* k^2 \Ft_L
\end{equation}
with domain $D(h^L) = \Ft_L^* D(k^2)$. In the case $L>0$, however we
must use the  unitary mapping $\hat{\Ft_L}$, defining
\begin{equation}
h^L = \hat{\Ft_L}^{-1} (k^2\oplus E_L) \hat{\Ft_L}
\end{equation}
with domain $D(h^L) = \hat{\Ft}_L^{-1}(D(k^2)\oplus\CC)$, where
$E_L\in \RR$ is arbitrary. We have restricted our operator to be
diagonal in the momentum representation in order to ensure that the
continuum eigenfunctions are still given by the $u_k(r)$. Note that
our construction does not uniquely determine the energy $E_L$ of the
bound state -- any $E_L$ results in an unbounded self-adjoint
operator whose continuum eigenfunctions have the required form. In
order to remove this ambiguity, we impose the additional requirement
of locality, in the form of a requirement that $h^L$ should agree
with $-d^2/dr^2$ on $C_0^\infty(0,\infty)$. Clearly this will
automatically restrict us to self-adjoint extensions of $-d^2/dr^2$
on this domain.

\begin{Prop} \label{Prop:loc}
$E_L=-|L|^{-2}$ is the unique value for which $h^L$ is local.
\end{Prop}
{\em Proof:} From Proposition~\ref{Prop:PW}, it follows that $\Ft_L
C_0^\infty(0,\infty)\subset D(k^2)$ and hence
$C_0^\infty(0,\infty)\subset D(h^L)$. For $f\in
C_0^\infty(0,\infty)$, we compute
\begin{eqnarray}
h^L f & =& \Ft_L^* k^2 \Ft_L f +
E_L\ket{\chi_L}\inner{\chi_L}{f} \nonumber \\
&=& -\Ft_L^*\Ft_L f^{\prime\prime}  +
E_L\ket{\chi_L}\inner{\chi_L}{f} \nonumber \\
&=& - f^{\prime\prime}
+\ket{\chi_L}(E_L\inner{\chi_L}{f}-
\inner{\chi_L}{f^{\prime\prime}}) \nonumber \\
&=& - f^{\prime\prime} +
(E_L+|L|^{-2})\ket{\chi_L}\inner{\chi_L}{f}
\end{eqnarray}
where we have used the fact that $k^2\Ft_L f = -\Ft_L f^{\prime\prime}$ for
$f\in
C_0^\infty(0,\infty)$ and also that $(-d^2/dr^2 |_{
C_0^\infty(0,\infty)})^*\chi_L=-|L|^{-2}\chi_L$. The result follows
immediately. $\square$

We now determine the domain of $h^L$ explicitly. This is, of course,
well known; we discuss it here only to show how it may be derived
within the terms of our construction.
\begin{Thm}
$h^L$ has domain
\begin{equation}
D(h^L) = \{ \varphi\mid \varphi,\varphi^\prime\in AC_{\rm
loc}(0,\infty);~\varphi^{\prime\prime}\in
L^2(0,\infty);~\varphi(0)+L\varphi^\prime(0) = 0\}
\end{equation}
where the boundary condition $\varphi(0)+L\varphi^\prime(0) = 0$ is
to be interpreted as $\varphi(0)=0$ for $L=0$, and
$\varphi^\prime(0)=0$ for $L=\infty$.
\end{Thm}
{\em Proof:} We give details for the case $L<0$ and indicate how the
proof is modified for the remaining cases.  Note that
$C_0^\infty(0,\infty)\subset\HH_k$ is a core for $k^2$, and hence
$\DD = \Ft_L^* C_0^\infty(0,\infty)$ is a core for $h^L$. Any
$f\in\DD$ may be written
\begin{equation}
f= (\St^{-1}-\Ct^{-1}kL)\varphi
\end{equation}
where $\varphi=(1+(kL)^2)^{-1/2}\Ft_Lf\in C_0^\infty(0,\infty)$. By
Proposition~\ref{Prop:PW} and identities of the sine and cosine
transforms, we find that $f(r)=g(r)-Lg^\prime(r)$, where
$g(r)=\St^{-1}\varphi$ is odd, entire analytic, and decreasing faster
than polynomially as $r\rightarrow \infty$. In particular, $f(0) +
Lf^\prime(0) = 0$ for all $f\in\DD$. Moreover, there is at least one
$f_0\in\DD$ for which $f_0(0)\not=0$, for otherwise we would have
$(\Ct^{-1} k(1+(kL)^2)^{-1/2}\eta)(0)=0$ for all $\eta$ in the dense
set $C_0^\infty(0,\infty)$, obtaining a contradiction. Now $h^L$
agrees with $-d^2/dr^2$ (acting in the sense of distributions) on
$\DD$, and is symmetric on this domain. Thus
\begin{equation}
D(h^L)=D((h^L|_{\DD})^*)=
\{\varphi\mid\inner{\varphi}{f^{\prime\prime}}
=\inner{\varphi^{\prime\prime}}{f},~\hbox{for all}~ f\in\DD\}
\end{equation}
and the required result follows easily. The cases $L=0,\infty$ may be
treated similarly, using $\St^{-1}C_0^\infty(0,\infty)$ and
$\Ct^{-1}C_0^\infty(0,\infty)$ as cores. For $L>0$, the appropriate
core is $\DD=\Ft_L^*
C_0^\infty(0,\infty)\oplus\{\lambda\chi_L(r)\mid\lambda\in\CC\}$.
$\square$

We can assemble the full PI Hamiltonian acting on $L^2(\RR^3)$: with
respect to the decomposition
\begin{equation}
L^2(\RR^3) = \bigoplus_{\ell=0}^\infty
L^2(\RR^+,r^2dr)\otimes {\cal K}_\ell    \label{eq:decmp}
\end{equation}
where ${\cal K}_\ell$ is the subspace of $L^2(\SS^2,d\Omega)$
spanned by $Y_{\ell,-\ell},\ldots,Y_{\ell,\ell}$, $H^L$ is defined by
\begin{equation}
H^L = U^*h^L U\otimes \II \oplus
\bigoplus_{\ell=1}^\infty U^*\bar{h}_\ell U\otimes\II
\end{equation}
where $U:L^2(\RR^+,dr)\rightarrow L^2(\RR^+,r^2dr)$ is the unitary
operator $(Uf)(r)=rf(r)$, and $\bar{h}_\ell$ ($\ell\ge 1$) is the
unique self-adjoint extension of $-d^2/dr^2+\ell(\ell+1)/r^2$ on
$C_0^\infty(0,\infty)\subset\HH_r$.

To summarise, we have seen how the usual class of point interactions
may be constructed from a consideration of the required scattering
theory.

\sect{Resonance Point Interactions}

We now construct the class of RPI models by a similar method to that
used in the previous section. In this case, the relevant heuristic
spectral representation is again given by~(\ref{eq:sph}), but with
$\delta_0(k)$ now specified by~(\ref{eq:RPIlow}) for some
$M\in\RR\cup\{\infty\}$. Accordingly, we consider the bounded mapping
$\Tt_M$ defined by
\begin{equation}
\Tt_M = (1+(kM)^2)^{-1/2}\Ct +
kM(1+(kM)^2)^{-1/2}\St.
\end{equation}
In the special cases $M=0,\infty$, $\Tt_M$ reduces to $\Ct$ and $\St$
respectively. Hence in these cases, $\Tt_M$ is unitary from $\HH_r$
to $\HH_k$ and the operators $\hr^M=\Tt_M^* k^2 \Tt_M$ are well
defined self-adjoint operators. Comparing with the previous section,
we see that $\hr^0=h^\infty$, and $\hr^\infty=h^0$.

We now consider the remaining cases. Firstly, we compute
$\Tt_M^*\Tt_M$. We have
\begin{eqnarray}
\Tt_M^*\Tt_M &=& \II + \Ct^{-1}\frac{1}{1+(kM)^2}\Ct -
\St^{-1}\frac{1}{1+(kM)^2}\St \nonumber \\
 & & + \Ct^{-1}\frac{kM}{1+(kM)^2}\St +
\St^{-1}\frac{kM}{1+(kM)^2}\Ct
\end{eqnarray}
and so, by the arguments used in the previous section,
\begin{equation}
\Tt_M^*\Tt_M =
\left\{\begin{array}{cl} \II + \ket{\chi_M}\bra{\chi_M} & M>0 \\
                                    \II & M < 0 \end{array}\right.
\end{equation}
which should be compared with equation~(\ref{eq:FstarF}).

Computing $\Tt_M\Tt_M^*$, we have
\begin{eqnarray}
\Tt_M\Tt_M^* & = &
\II + \frac{1}{(1+(kM)^2)^{1/2}}
\Ct\St^{-1}\frac{kM}{(1+(kM)^2)^{1/2}}\nonumber \\
&&+ \frac{kM}{(1+(kM)^2)^{1/2}}
\St\Ct^{-1}\frac{1}{(1+(kM)^2)^{1/2}}
\end{eqnarray}
On the domain $\DD^M_+$ defined in~(\ref{eq:Dpm}), the last term may
be re-written:
\begin{eqnarray}
\frac{kM}{(1+(kM)^2)^{1/2}}
\St\Ct^{-1}\frac{1}{(1+(kM)^2)^{1/2}} & =&
\frac{M}{(1+(kM)^2)^{1/2}}
\Ct\frac{d}{dr}\Ct^{-1}\frac{1}{(1+(kM)^2)^{1/2}}\nonumber \\
 &= & \frac{-M}{(1+(kM)^2)^{1/2}}
\Ct\St^{-1}\frac{k}{(1+(kM)^2)^{1/2}}
\end{eqnarray}
Proposition~\ref{Prop:dense} shows that $\DD_+\oplus
\{\lambda\psi_M\mid \lambda\in\CC\}$ is dense in $\HH_k$, so we
require the action of $\Tt_M\Tt_M^*$ on $\psi_M$. Explicit
computation using the standard results
\begin{eqnarray}
(\Ct \chi_M)(k) =
\left(\frac{4|M|}{\pi}\right)^{1/2}\frac{1}{1+(kM)^2}; & &
(\St \chi_M)(k) =
\left(\frac{4|M|}{\pi}\right)^{1/2}\frac{k|M|}{1+(kM)^2}
\end{eqnarray} yields
\begin{equation}
(\Tt_M\chi_M)(k) =
\left\{\begin{array}{cl}
\sqrt{2}\psi_M(k)\frac{1-(kM)^2}{1+(kM)^2} & M<0 \\
\sqrt{2}\psi_M(k) & M>0
\end{array}\right.
\end{equation}
and
\begin{equation}
(\Tt_M^*\psi_M)(r) = \frac{1}{\sqrt{2}}(1 + {\rm sgn} M)\chi_M(r).
\end{equation}
from which we obtain
\begin{equation}
\Tt_M\Tt_M^*\psi_M = \left\{
\begin{array}{cl} 2\psi_M & M>0 \\ 0 & M<0. \end{array}
\right.
\end{equation}
Assembling our results, we have
\begin{equation}
\Tt_M\Tt_M^* =
\left\{\begin{array}{cl} \II+\ket{\psi_M}\bra{\psi_M} & M>0 \\
                                        \II-\ket{\psi_M}\bra{\psi_M} & M<0. \\
                      \end{array}\right.
\end{equation}

We now modify $\Tt_M$ in order to obtain a unitary operator. First
note that the $M<0$ case is analagous to the $L>0$ case for $\Ft_L$,
except that here, it is $\Tt_M\Tt_M^*$ rather than $\Ft_L^*\Ft_L$
which fails to be the identity. Accordingly, we extend the {\em
position} Hilbert space to $\HH_r\oplus\CC$ (with the obvious inner
product) and define the mapping $\hat{\Tt}_M$ by
\begin{eqnarray}
\hat{\Tt}_M : \HH_r\oplus \CC &\longrightarrow & \HH_k \nonumber \\
f\oplus \alpha & \longrightarrow & \Tt_M f + \alpha\ket{\psi_M}.
\end{eqnarray}
$\hat{\Tt}_M$ is unique up to re-phasing of $\psi_M$. We then have
\begin{Prop}
For $M<0$, $\hat{\Tt}_M$ is a unitary mapping from $\HH_r\oplus\CC$
to $\HH_k$ with inverse
\begin{equation}
\hat{\Tt}_M^{-1} f = \left[\Tt_M^*,\bra{\psi_M}\right]f = \Tt_M^*f
\oplus \inner{\psi_M}{f}.
\end{equation}
\end{Prop}
We may therefore construct a
self-adjoint operator $\hr^M$ on $\HH_r\oplus\CC$ by
\begin{equation}
\hr^M = \hat{\Tt}_M^{-1} k^2 \hat{\Tt}_M
\end{equation}
with domain $D(\hr^M)=\hat{\Tt}_M^{-1} D(k^2)$. We will return to the
physical interpretation of the `extra dimension' in Section 6.

We now examine the properties of $\hr^M$. By the same arguments as in
the case of $L>0$ PI, the domain of $\hr^M$ includes the (non-dense)
subspace  $\DD = \{ \phi\oplus 0 \mid \phi\in
C_0^\infty(0,\infty)\}$. Moreover, one can easily see that
\begin{eqnarray}
\hr^M(\phi\oplus 0) &=& \Tt_M^* k^2 \Tt_M \phi \oplus
\inner{\psi_M}{k^2\Tt_M\phi}
\nonumber \\
&= & -\Tt_M^* \Tt_M \phi^{\prime\prime} \oplus
\inner{\psi_M}{-\Tt_M\phi^{\prime\prime}}
\nonumber \\
&= & -\phi^{\prime\prime} \oplus 0
\end{eqnarray}
for $\phi\in C_0^\infty(0,\infty)$ where we
have used the fact that $\Tt_M^*\psi_M =0$ for $M<0$.
Hence, $\hr^M$ is a self-adjoint extension of the non-densely
defined operator
\begin{equation}
-\frac{d^2}{dr^2} \oplus 0 \quad
{\rm on}~\DD\subset L^2((0,\infty),dr)\oplus\CC
\end{equation}
The $\hr^M$ are therefore local, and belong to the class of models
considered by Pavlov \cite{Pav1} and also by Shondin \cite{Shond1}.
In the nomenclature of \cite{Shond1}, the $\hr^M$ for $M<0$ form a
subset of models of `type $B_2$'.

To describe the domain of $\hr^M$ explicitly, we use a similar
argument to that employed in the PI case, identifying
$\DD=\hat{\Tt}_M^{-1}C_0^\infty(0,\infty)$ as a core for $\hr^M$.  If
$g(k)\in C_0^\infty(0,\infty)$, then
$\inner{\psi_M}{g}=|M|^{1/2}(\Tt_M^*g)(0)$, and so $\DD$ may be
written
\begin{equation}
\DD = \{ f\oplus |M|^{1/2}f(0)\mid f\in
\Tt_M^*C_0^\infty(0,\infty)\}.
\end{equation}
Any function $f\in\Tt_M^*C_0^\infty(0,\infty)$ may be written
$f(r)=g(r)-M g^\prime(r)$, where $g(r)$ is analytic, even, and
decreasing faster than polynomially as $r\rightarrow 0$. As a
consequence, $f^\prime(0)=-Mf^{\prime\prime}(0)$. On $\DD$, $\hr^M$
has action
\begin{eqnarray}
\hr^M f\oplus |M|^{1/2}f(0) &=&
 -f^{\prime\prime}\oplus -|M|^{1/2}f^{\prime\prime}(0) \nonumber \\
& = & -f^{\prime\prime}\oplus -|M|^{-1/2}f^{\prime}(0) \label{eq:act}
\end{eqnarray}
The domain of $\hr^M$ is equal to $D((\hr^M |_{\DD})^*)$.
We have
\begin{eqnarray}
\inner{\varphi\oplus\Phi_1}{\hr^M (f\oplus |M|^{1/2}f(0))} &=&
\inner{-\varphi^{\prime\prime}}{f} + \overline{\varphi(0)}f^\prime(0)
 -\overline{\varphi^\prime(0)}f(0) \nonumber\\
& & -\overline{\Phi}_1|M|^{-1/2}f^\prime(0)
\end{eqnarray}
for any $\varphi$ such that $\varphi^{\prime\prime}\in L^2(0,\infty)$
in the sense of distributions. Note that $-d^2/dr^2$ is {\em not}
symmetric on $\Tt_M^* C_0^\infty(0,\infty)$.
 Hence, if $\varphi\oplus\Phi_1\in D(\hr^M)$, with $\hr^M
\varphi\oplus\Phi_1=-\varphi^{\prime\prime}\oplus\Phi_2$, then
\begin{equation}
 \overline{\varphi^\prime(0)}f(0) -\overline{\varphi(0)}f^\prime(0)
= \overline{\Phi}_1|M|^{-1/2}f^\prime(0)+\overline{\Phi}_2|M|^{1/2}
f(0).
\end{equation}
for all $f\in \Tt_M^*C_0^\infty(0,\infty)$. Moreover, $f(0)$ and
$f^\prime(0)$ are independent on this domain. The following is then
immediate.  \begin{Thm}
The domain of $\hr^M$ is
\begin{equation}
D(\hr^M) = \{\varphi\oplus\Phi\mid \varphi,\varphi^\prime\in
AC_{\rm
loc}(0,\infty);~\varphi^{\prime\prime}\in L^2(0,\infty);~
\Phi=|M|^{1/2}\varphi(0)\}
\label{eq:domRPI}
\end{equation}
with action
\begin{equation}
\hr^M (\varphi\oplus |M|^{1/2}\varphi(0)) =
-\varphi^{\prime\prime}\oplus -|M|^{-1/2}\varphi^{\prime}(0).
\end{equation}
\end{Thm}
In contrast to the usual PI case, the boundary condition corresponding
to~(\ref{eq:domRPI}) is energy dependent: the eigenfunction equation
at energy $k^2$ is equivalent to the  equation
$-\varphi^{\prime\prime} = k^2\varphi$, with boundary condition
\begin{equation}
k^2M\varphi(0) = \varphi^\prime(0)
\end{equation}
and the generalised eigenfunction at energy $k^2$ is therefore
\begin{equation}
u_k = \left(\frac{2}{\pi}\right)^{1/2}\sin (kr+\delta_0(k)) \oplus
|M|^{1/2}\sin\delta_0(k)
\end{equation}
with $\delta_0(k)$ given by~(\ref{eq:RPIlow}).

As a result of our construction, the spectral properties of $\hr^M$
are easily identified: $\hr^M$ has purely absolutely continuous
spectrum $\sigma(\hr^M)=\sigma_{\rm ac}(\hr^M)=[0,\infty)$.

Turning to the case $M>0$, we note that our original choice of
continuum eigenfunctions form an `over complete' set. We can remedy
this by extending to a Pontryagin space, the heuristic motivation
being that this will allow us to subtract off the `excess
probability'. It is only possible to construct unitary mappings
between Pontryagin spaces with the same rank of indefiniteness, so we
must extend both position and  momentum spaces to Pontryagin spaces.
As before, we find that the failure of unitarity is located in a one
dimensional subspace, which suggests that we choose Pontryagin spaces
of form $\Pi=L^2(0,\infty)\ominus\CC$, where the indefinite inner
product is  given by
\begin{equation}
\inner{f\ominus \alpha}{g\ominus\beta}_\Pi = \inner{f}{g}_{L^2} -
\overline{\alpha}\beta .
\end{equation}

We define the mapping $\hat{\Tt}_M$, unique up to re-phasing of
$\psi_M$ and $\chi_M$, by
\begin{eqnarray}
\hat{\Tt}_M : \HH_r\ominus \CC &\longrightarrow & \HH_k\ominus\CC
\nonumber \\
f\ominus \alpha & \longrightarrow & (\Tt_M f -
\alpha\ket{\psi_M})\ominus
(-\inner{\chi_M}{f}_{L^2} + \alpha\sqrt{2}) .
\end{eqnarray}
A short computation proves
\begin{Prop}
For $M>0$, $\hat{\Tt}_M$ is a unitary mapping from $\HH_r\ominus \CC$ to
$\HH_k\ominus\CC$, with inverse
\begin{eqnarray}
\hat{\Tt}_M^{-1} : \HH_k\ominus \CC &\longrightarrow &
\HH_r\ominus\CC
\nonumber \\
f\ominus \alpha & \longrightarrow & (\Tt_M^* f +
\alpha\ket{\chi_M})\ominus
(\inner{\psi_M}{f}_{L^2} + \alpha\sqrt{2})
\end{eqnarray}
\end{Prop}

Hence, we can construct the RPI Hamiltonian for $M>0$ on
$L^2((0,\infty),dr)\ominus\CC$:
\begin{equation}
\hr^M = \hat{\Tt}_M^{-1} \left(k^2 \ominus E_M \right)\hat{\Tt}_M
\end{equation}
(We have again chosen the Hamiltonian to be diagonalised by
$\hat{\Tt}_M$.) As in the case of PI for $L>0$, a single free real
parameter $E_M$ is introduced by our procedure.  $E_M$ is interpreted
as the energy of a negative-normed eigenstate
$\ket{\chi_M}\ominus\sqrt{2}$. Such a state is undesirable in a
quantum theory. Accordingly, we decompose the position Pontryagin
space $\Pi_r$ into the orthogonal subspaces $\HH_+$ and $\LL$ (which
are independent of the choice of $E_M$) given by
\begin{eqnarray}
\Pi_r &=& \HH_+\domin \LL  \nonumber\\
\HH_+ &=& \hat{\Tt}_M^{-1}(\HH_k\ominus 0) \nonumber \\
\LL &=& \hat{\Tt}_M^{-1}(0\ominus\CC)
=\{\lambda (\ket{\chi_M}\ominus\sqrt{2})\mid
\lambda\in\CC\}
\end{eqnarray}
where we use the notation $\domin$ in contrast to the
decomposition $\Pi_r = \HH_r\ominus\CC$.

Due to the diagonal structure of $\hr^M$ in the momentum
representation, $\hr^M$ respects this decomposition
\begin{equation}
\hr^M = \hrp^M\domin E_M. \label{eq:dc}
\end{equation}
Moreover, $\HH_+$ is an intrinsically complete, positive definite
subspace of $\Pi_r$ of unit co-dimension, orthogonal to the
negative-normed eigenstate found above; it may therefore naturally be
identified as the space of physical states. The operator $\hrp^M$ and
the space $\HH_+$ are independent of the value of $E_M$, so to some
extent this parameter has no physical meaning. However, if one wishes
the operator $\hr^M$ to be local on the Pontryagin space $\Pi_r$, one
can identify the value $E_M=-|M|^{-2}$ as the unique value compatible
with this requirement by similar arguments to those used in the PI
case for $L>0$.

We now describe the domain of $\hrp^M$ in more detail. We have
$D(\hrp^M)=\hat{\Tt}_M^{-1} (D(k^2)\ominus 0)\subset\HH_+$. The
domain $\DD= \hat{\Tt}_M^{-1}(C_0^\infty(0,\infty)\ominus 0)$ is a
core for $\hrp^M$, and may be written
\begin{equation}
\DD = \{ f\ominus |M|^{1/2}f(0)\mid f\in \Tt_M^*C_0^\infty(0,\infty)\}.
\end{equation}
As before, we note that any function
$f\in\Tt_M^*C_0^\infty(0,\infty)$ may be written $f(r)=g(r)-M
g^\prime(r)$, where $g(r)$ is analytic, even, and decreasing faster
than polynomially as $r\rightarrow 0$, and that
$f^\prime(0)=-Mf^{\prime\prime}(0)$ in consequence. The action of
$\hrp^M$ on $\DD$ is given by
\begin{equation}
\hrp^M (f\ominus
|M|^{1/2}f(0)) = -f^{\prime\prime}\ominus  |M|^{-1/2}f^\prime(0)
\end{equation}
An exact analogue of the argument used in the $M<0$ case then yields
\begin{Thm}
The domain of $\hrp^M$ is given by
\begin{equation}
D(\hrp^M) =
\{\varphi\ominus\Phi\mid \varphi,\varphi^\prime\in AC_{\rm
loc}(0,\infty);~\varphi^{\prime\prime}\in L^2(0,\infty);~
\Phi=|M|^{1/2}\varphi(0)\}
\end{equation}
with action
\begin{equation}
\hrp^M (\varphi\ominus\Phi) = -\varphi^{\prime\prime}\ominus
|M|^{-1/2}\varphi^{\prime}(0).
\end{equation}
\end{Thm}
We therefore have the full domain $D(\hr^M)=D(\hrp^M)\domin\LL$
\begin{eqnarray}
D(\hr^M) &=& \{(\varphi\ominus\Phi)+
\alpha(\ket{\chi_M}\ominus\sqrt{2})\mid
\varphi,\varphi^\prime\in AC_{\rm loc}(0,\infty); \nonumber\\
&&\qquad\qquad\varphi^{\prime\prime}\in L^2(0,\infty);~
\Phi=|M|^{1/2}\varphi(0);~\alpha\in\CC\}
\end{eqnarray}
and action
\begin{equation}
\hr^M (\varphi\ominus\Phi)
+\alpha(\ket{\chi_M}\ominus\sqrt{2}) =
-\varphi^{\prime\prime}\ominus
|M|^{-1/2}\varphi^{\prime}(0) +
\alpha E_M(\ket{\chi_M}\ominus\sqrt{2}).
\end{equation}

As before, the $\hr^M$ exhibit energy dependent boundary conditions:
solving the eigenvalue equation at energy $k^2$ yields
solutions obeying
\begin{equation}
k^2M\varphi(0)=\varphi^\prime(0).
\end{equation}

The spectral properties of $\hr^M$ are as follows:
$\sigma(\hr^M)=\sigma_{\rm pp}(\hr^M)\cup\sigma_{\rm ac}(\hr^M)$,
with $\sigma_{\rm pp}(\hr^M)=\{E_M\}$ and $\sigma_{\rm
ac}(\hr^M)=[0,\infty)$.

We can assemble the full RPI models on $\RR^3$ as before: defining
$\Hr^M$ by
\begin{equation}
\Hr^M = \widetilde{U}^*\hr^M \widetilde{U}\otimes \II \oplus
\bigoplus_{\ell=1}^\infty U^*\bar{h}_\ell U\otimes \II
\end{equation}
where, for $M<0$, $\widetilde{U}(f\oplus \alpha)=rf(r)\oplus\alpha$,
and for $M>0$, $\widetilde{U}(f\ominus \alpha)=rf(r)\ominus\alpha$.

\sect{Scattering Theory}

The original aim of our construction was to produce PI and GPI
Hamiltonians with a given $S$-wave phase shift. It is therefore
expedient to check that the models described above actually exhibit
the required behaviour. The usual method of demonstrating the
existence and completeness of M{\o}ller wave operators for PI and GPI
models is to show that their resolvents are trace class perturbations
of that for the free Laplacian, and then to apply Kuroda-Birman
theory \cite{RSiii}. Instead, we will use a method which builds on
our construction and yields the M{\o}ller operators (and their
completeness) directly.

We work in the $S$-wave only, and employ a two space setting: let $B$
be self-adjoint on $\HH_1$, $A$ be self-adjoint on $\HH_2$ and $J$ be
a bounded operator from $\HH_1$ to $\HH_2$. Then the M{\o}ller
operators $\Omega^\pm(A,B;J)$ are defined by
\begin{equation}
\Omega^\pm(A,B:J) = \lim_{t\rightarrow\mp\infty}
e^{iAt}Je^{-iBt}P_{\rm ac}(B)
\end{equation}
and are said to be complete if the closure of ${\rm
Ran}\Omega^\pm(A,B;J)$ is equal to ${\rm Ran} P_{\rm ac} A$.

Our results in this section follow from
\begin{Prop} \label{Prop:scat}
$U_{-t}\Ct\St^{-1}U_t\rightarrow \pm i\II$ as
$t\rightarrow\mp\infty$, where $U_t$ is
multiplication by $e^{-ik^2t}$ on $\HH_k$.
\end{Prop}
{\em Proof:} For any $u(k)\in
C_0^\infty(0,\infty)$, we compute
\begin{eqnarray}
\| U_{-t}\Ct\St^{-1}U_{t}u(k)\mp i u(k)\|^2 & = &
\|(\Ct^{-1}\pm i\St^{-1})U_t u(k)\|^2 \nonumber\\
& = & \frac{2}{\pi}\int_0^\infty dr
\left|\int_0^\infty dk e^{i(\pm kr-k^2t)} u(k)\right|^2
\end{eqnarray}
and note that the last expression vanishes if $t\rightarrow\mp\infty$
by (non)-stationary phase arguments (see the Corollary to Theorem
XI.14 in \cite{RSiii}). $\square$

Using this result, we see that if
$\Qt=\cos\delta_0(k)\St+\sin\delta_0(k)\Ct$, then
\begin{equation}
U_{-t}\Qt\St^{-1}U_t\rightarrow e^{\pm i\delta_0(k)}
\end{equation}
as $t\rightarrow\mp\infty$. It is then easy to construct the
M{\o}ller wave operators for the PI and RPI models. For the PI case,
we take $\HH_1=\HH_2$ and $J$ to be the identity, writing
$\Omega^\pm(A,B)$ for the wave operators.

\begin{Thm} $\Omega^\pm(h^L,h^0)$ exist, are complete, and given by
\begin{equation}
\Omega^\pm(h^L,h^0)=
\Ft_L^*e^{\pm i\delta_0(k)}\St \label{eq:fac}
\end{equation}
where $\delta_0(k)$ is given by~(\ref{eq:PIlow}).
\end{Thm}
{\em Proof:} For $L\le 0$, $L=\infty$, the existence and form of the
M{\o}ller operators is immediate from the above, and the definition
of $h^L$ as $\Ft_L^*k^2\Ft_L$. Completeness holds because all three
factors in~(\ref{eq:fac}) are unitary, and hence ${\rm Ran}
\Omega^\pm(h^L,h^0)=\HH_r={\rm Ran} P_{\rm ac}h^L$. For $L>0$, we
have
\begin{equation}
e^{ih^Lt}e^{-ih^0t} =
\Ft_L^*U_{-t}\Ft_L\St^{-1}U_t\St +
e^{-i|L|^{-2}t}\ket{\chi_L}\bra{\chi_L}\St^{-1}U_t\St
\end{equation}
The second term vanishes as $|t|\rightarrow\infty$ by another
non-stationary phase argument, and the required result follows
because ${\rm Ran} \Ft_L^*={\rm Ran}P_{\rm ac} h^L$. $\square$

\begin{Thm} For $M<0$, $\Omega^\pm(\hr^M,h^0;J)$ exist, are complete,
and given by
\begin{equation}
\Omega^\pm(\hr^M,h^0)=
\hat{\Tt}_M^{-1}e^{\pm i\delta_0(k)}\St
\end{equation}
where $\delta_0(k)$ is given by~(\ref{eq:RPIlow}), and
$J:\HH_r\rightarrow\HH_r\oplus\CC$ is defined by $Jf=f\oplus 0$.
\end{Thm}
{\em Proof:} The existence and form of the operators is due to
Proposition~\ref{Prop:scat} and the following observation along with
the form of $\Tt_M$ and the fact that $\hat{\Tt_M}J=\Tt_M\oplus 0$.
Completeness holds because ${\rm Ran}\hat{\Tt}_M^{-1}=\HH_r={\rm Ran}
P_{\rm ac}\hr^M$. $\square$

For the case $M>0$, our two spaces are $\HH_r$ and $\HH_+$, the
physical state space.
\begin{Thm}
For $M>0$,
$\Omega^\pm(\hr^M,h^0;J)$ exist, are complete, and given by
\begin{equation} \Omega^\pm(\hr^M,h^0)=
\hat{\Tt}_M^{-1}e^{\pm i\delta_0(k)}\St
\end{equation}
where $\delta_0(k)$ is given by~(\ref{eq:RPIlow}), and
$J:\HH_r\rightarrow\HH_r\ominus\CC$ is defined by
$Jf=P_+(f\ominus 0)$, with $P_+$ the
orthogonal projector onto $\HH_+$.
\end{Thm}
{\em Proof:} The argument proceeds as before, completeness holding
because the wave operators are isometries from $\HH_r$ to
$\HH_+=P_{\rm ac}\hr^M$. $\square$

We conclude that our construction does indeed yield the required
scattering theory, and also (because of the way in which the PI and
RPI Hamiltonians were defined) that complete M{\o}ller operators may
easily and explicitly be determined.

\sect{Physical Interpretation}

In this section, we show how RPI models may be used to model
Schr\"{o}dinger operators $H=-\triangle+V$, where $V$ is smooth,
spherically symmetric, compactly supported within radius $a$ of the
origin, and exhibits a zero energy resonance. Our methodology is
analagous to that developed in \cite{KF}, in which the non-resonant
case is discussed. Here, we  develop a formalism (the fitting
formula) for selecting the `best fit' RPI for such operators. The
range of energies for which the approximation is valid can be
determined by a `believability' analysis analagous to that developed
in \cite{KF}. We will not do this here.

The $S$-wave wavefunction at wavenumber $k$ obeys
\begin{equation}
\left\{-\frac{d^2}{dr^2} +V(r)\right\} u_k(r) =
k^2 u_k(r), \label{eq:Seq}
\end{equation}
where $u(r)$ obeys regular boundary conditions, i.e. $u(0)=0$.
Because $V$ is supported within radius $a$ of the origin, the phase
shift is given by
\begin{equation}
\cot\delta_0(k)=\frac{ka \sin ka + D(k) \cos ka}{ka\cos ka -
D(k)\sin ka},
\label{eq:ddf}
\end{equation}
where $D(k)=au_k^\prime(a)/u_k(a)$. Expanding D(k) in powers of
$(ka)^2$, $D(k) = D_0+D_1 (ka)^2 + O((ka)^4)$, and substituting
in~(\ref{eq:ddf}), we find
\begin{equation}
\cot\delta_0(k) = \frac{D_0 + (1+D_1-D_0)(ka)^2+O((ka)^4)}{(1-D_0)ka
+O((ka)^3)}.
\end{equation}
This then leads to the expression $L=a(D_0-1)/D_0$ for the scattering
length. Clearly, a zero energy resonance (infinite scattering length)
occurs when $D_0 =0$. In this case, the leading order behaviour of
$\delta_0(k)$ is given by
\begin{equation}
\cot\delta_0(k) = (1+D_1)ka +O((ka)^3).
\end{equation}
Comparing with~(\ref{eq:RPIlow}), we find that the leading order
approximation to the dynamics is given by the RPI $\Hr^M$, with
$M=a(1+D_1)$. Thus it suffices to compute $D_1$ for the potential of
interest. To do this, we use the well-known formula (see, e.g.
\cite{Merz}, p.236)
\begin{equation}
D(k_1)-D(k_2) = -((k_1a)^2-(k_2a)^2)
\frac{\int_0^a u_{k_1}(r)u_{k_2}(r) dr}{au_{k_1}(a)u_{k_2}(a)}
\end{equation}
Using this to expand $D(k)$ about $k=0$, we find
\begin{equation}
D_1 = -\frac{a^{-1}\int_0^a u_0(r)^2 dr}{u_0(a)^2}
\end{equation}
and thus arrive at the {\em fitting formula} (cf. \cite{KF}
\begin{equation}
M= a \left( 1 -\frac{a^{-1}\int_0^a u_0(r)^2 dr}{u_0(a)^2}\right)
\label{eq:fit}
\end{equation}
The best fit RPI can therefore be computed in terms of the zero
energy solution to~(\ref{eq:Seq}). In addition, the labelling
parameter obeys the bound
\begin{equation}
-\infty \le M < a.
\end{equation}
Moreover, this bound is best possible: for any $M$ in the above
range, one can clearly find a smooth function $u_0(r)$ satisfying
regular boundary conditions at the origin, $u_0$ constant for $r>a$
and such that~(\ref{eq:fit}) holds. Then the potential defined by
$V(r)=u_0^{\prime\prime}(r)/u_0(r)$ has infinite scattering length,
and scattering theory approximated at leading order by $\Hr^M$.

It is interesting to note that the RPI models themselves
obey~(\ref{eq:fit}) in the following sense: the zero energy
generalised eigenfunction for $M<0$ is given by
\begin{equation}
u_0 = 1 \oplus |M|^{1/2}.
\end{equation}
Interpreting the integral in~(\ref{eq:fit}) for elements
$f\oplus\alpha$ of $\HH_r\oplus\CC$ as $\int_0^a |f(r)|^2 dr +
|\alpha|^2$, the right hand side of equation~(\ref{eq:fit}) is then
equal to $M$. Similarly, for $M>0$, we interpret the integral for
$f\ominus\alpha\in\HH_r\ominus \CC$ as $\int_0^a |f(r)|^2 dr -
|\alpha|^2$ and use the generalised zero energy eigenfunction
\begin{equation}
u_0 = 1\ominus |M|^{1/2}
\end{equation}
to yield the value $M$.

We briefly consider the interpretation of the extension to the
Hilbert space $\HH_r$ in the case $M<0$. From the fitting formula, it
is clear that $M$ is negative if and only  if the mean square value
of $u_0(r)$ within $r<a$ exceeds $u_0(a)^2$. Hence this case is used
to model Schr\"{o}dinger operators whose zero energy generalised
eigenfunctions are peaked inside the interaction region. This is
characteristic of resonant behaviour and corresponds to a physical
picture of a particle being detained inside the region, before being
gradually released. If the radius of support is shrunk to a point,
the particle must be completely removed from the space in order to
model this process. It is therefore natural that the Hilbert space be
extended in this case.

One may study the time behaviour of the proportion of a wavepacket in
the extra dimension. Starting with a normalised vector $0\oplus 1$,
the evolved packet is given by $\hat{\Tt}_M^{-1}e^{-ik^2t}\psi_M$.
The proportion remaining in the initial state at time $t$ is
\begin{eqnarray}
\varphi(t) &=& \inner{\psi_M}{e^{-ik^2t}\psi_M} \nonumber \\
&=& \frac{2|M|}{\pi}\int_0^\infty dk \frac{e^{-ik^2t}}{1+(kM)^2}
\end{eqnarray}
which may be approximated using the method of stationary phase to
give
\begin{equation}
\varphi(t) \sim \frac{2|M|}{\sqrt{\pi t}} e^{-i\pi/4} + O(t^{-1})
\end{equation}
as $t\rightarrow \infty$. It is interesting that the decay is not
exponential. The $M>0$ case does not admit such a simple
interpretation due to the redefinition of the space of physical
states.

As a simple example of the use of the fitting formula, we consider
the example of a square well with a zero energy resonance. Setting
$V(r)=-((n+\frac{1}{2})\pi/a)^2$ ($n=0,1,2,\ldots$) for $r<a$ and
$V(r)=0$ for $r>a$, we find $u_0(r)=\sin (n+\frac{1}{2})\pi r/a$.
Equation~(\ref{eq:fit}) then yields $M=a/2$ for any value of $n$.

\sect{Conclusion}

We have seen how PI and GPI models may be constructed and studied
using integral transforms suggested by the scattering data. In
addition, we have constructed a new class of RPI models (the case
$M>0$). The RPI models obey energy dependent boundary conditions and
are defined on a Hilbert or Pontryagin space which extends the usual
space of states. In the Pontryagin case a positive definite invariant
subspace may be constructed which may be interpreted as the physical
state space.

It is instructive to compare our methods with those usual in
constructions of GPI models on Pontryagin spaces \cite{Shond2}. One
starts with the desire to include wavefunctions with distributional
terms in the space of states. Accordingly, choosing a dense subspace
$\DD\subset\HH_r$ and an anti-linear functional $\omega$ (more
generally $m$ such functionals) on $\DD$ ($\omega\not\in\HH_r$) one
constructs a linear space $\HH$ generated by elements
\begin{equation}
\psi=\psi_0+\lambda\omega
\end{equation}
with $\psi_0\in\DD$, $\lambda\in\CC$. The next step is to extend the
Hilbert space inner product $\inner{\cdot}{\cdot}$ from $\DD$ to a
sesquilinear form $(\cdot,\cdot)$ on $\HH$ by defining
\begin{eqnarray}
(\psi_0,\varphi_0) = \inner{\psi_0}{\varphi_0} & &
\psi_0,\varphi_0\in\DD \nonumber \\
(\psi_0,\omega ) = \omega(\psi_0) & & \psi_0\in\DD
\end{eqnarray}
(recall that $\omega$ is anti-linear). This leaves $(\omega,\omega)$
as a free parameter which is usually fixed on the basis of a
physically inspired renormalisation of certain divergent integrals.
The form $(\cdot,\cdot)$ is in general indefinite; however, there is
a closely related positive definite inner product which is used to
form the completion $\widetilde{\HH}$ of $\HH$. With respect to
$(\cdot,\cdot)$, $\widetilde{\HH}$ is a Pontryagin space
$\HH_r\oplus\CC\ominus\CC$. One can then seek the GPI models which
`live' on $\widetilde{\HH}$.

The present treatment has various advantages over this procedure.
Firstly, one need not determine the distribution $\omega$ in advance;
moreover the form of the Pontryagin space is suggested naturally by
the over completeness of the generalised
eigenfunctions~(\ref{eq:gefn}). Indeed, distributions need never
explicitly occur in our construction; although it is clear (e.g. from
the domain of definition of $\hr^M$) that the component of the
wavefunction in the `extra dimension' carries a distributional
interpretation. Finally, it is not clear that the usual procedure
can encompass the $M>0$ RPI models without modification, as the
natural Pontryagin space is $\HH_r\ominus\CC$ rather than
$\HH_r\oplus\CC\ominus\CC$. It is possible that these models are
restrictions of GPI models in the larger space; however, it seems
more likely to us that a better starting point would be to choose a
non-dense test space $\DD$ consisting of vectors orthogonal to a
1-dimensional subspace. It is not presently clear exactly how this
would be implemented, nor what the appropriate distribution $\omega$
would be.

The construction presented here has so far only been employed for
interactions whose scattering is non-trivial only in the $S$-wave.
One of the great strengths of the usual constructions is that the
generalisation to higher angular momenta is relatively
straightforward. It would be interesting to extend the present
treatment to this case.

Finally, it is of interest to understand whether RPI models can arise
as limits of sequences of Schr\"{o}dinger operators with potentials
of compact support decreasing to the origin, e.g. in the spirit of
\cite{KF} in which the usual class of PI is exhibited as strong
resolvent limits of such sequences (see \cite{Alb} for a treatment
using sequences of scaled potentials in the norm resolvent topology).
We hope to address these issues elsewhere.

{\em Acknowledgements:} The notions of fitting formulae and the
general methodology of Section 6 are due in origin to Bernard Kay
\cite{KF}. I also thank Bernard for useful conversations and
encouragement, and Churchill College, Cambridge, for financial
support under a Gateway Studentship.

\end{document}